\documentclass[manyauthors,nocleardouble,COMPASS]{cernphprep}
\usepackage{bm}
\usepackage{color}

\usepackage{amssymb}
\usepackage{amsmath}
\usepackage{amsbsy}
\usepackage{times}
\usepackage{epsfig} 
\usepackage{colordvi}
\usepackage{graphicx}
\usepackage{cite}
\usepackage{hyperref}
\usepackage{multicol}
\usepackage{booktabs}
\usepackage{ulem}
\usepackage{lineno}
\usepackage{lscape}
\usepackage{xcolor}
\usepackage{orcidlink}

\RequirePackage[T1]{fontenc}
\definecolor{violet}{rgb}{0.2,0.0,0.8}

\begin{document}

%\linenumbers

\begin{titlepage}

\PHnumber{2025-291}
\PHdate{\today}  
\title{Addendum to multiplicities of charged pions, kaons and unidentified charged hadrons on an isoscalar target  measured by COMPASS Collaboration }

\date{}
\Collaboration{The COMPASS Collaboration}
\ShortAuthor{The COMPASS Collaboration}

\begin{abstract} \label{abstract}

The COMPASS Collaboration has recently published an article ``Multiplicities of positive and negative pions, kaons, and unidentified hadrons from deep-inelastic scattering of muons off a liquid hydrogen target'', Phys. Rev. D 112 (2025) 012002. In contrast to earlier COMPASS publications on similar topics, the aforementioned article features an enhanced treatment of QED radiative corrections, employing the DJANGOH Monte Carlo generator. This methodological improvement led to corrections that are up to 12\% larger in the low-$x$, high-$z$ region compared to the previously applied ones.
To ensure consistent treatment of COMPASS data sets obtained using both isoscalar and proton targets, this paper presents an updated set of isoscalar multiplicities based on DJANGOH-derived radiative corrections. The present results supersede those published in Phys. Lett. B 764 (2017) 1 and Phys. Lett. B 767 (2017) 133.

\end{abstract}

\vspace{2cm}

\vfill
\Submitted{(to be submitted to PLB)}

\end{titlepage}

{\pagestyle{empty}  
\center{\textbf{The COMPASS Collaboration}}

\vspace{10pt}
\begin{flushleft}
G.~D.~Alexeev$^\textrm{{\footnotesize\hyperlink{hl:dubna}{28}}}$\orcidlink{0009-0007-0196-8178},
M.~G.~Alexeev$^\textrm{{\footnotesize\hyperlink{hl:turin_u}{20},\hyperlink{hl:turin_i}{19}}}$\orcidlink{0000-0002-7306-8255},
C.~Alice$^\textrm{{\footnotesize\hyperlink{hl:turin_u}{20},\hyperlink{hl:turin_i}{19}}}$\orcidlink{0000-0001-6297-9857},
A.~Amoroso$^\textrm{{\footnotesize\hyperlink{hl:turin_u}{20},\hyperlink{hl:turin_i}{19}}}$\orcidlink{0000-0002-3095-8610},
V.~Andrieux$^\textrm{{\footnotesize\hyperlink{hl:illinois}{33}}}$\orcidlink{0000-0001-9957-9910},
V.~Anosov$^\textrm{{\footnotesize\hyperlink{hl:dubna}{28}}}$\orcidlink{0009-0003-3595-9561},
K.~Augsten$^\textrm{{\footnotesize\hyperlink{hl:praguectu}{4}}}$\orcidlink{0000-0001-8324-0576},
W.~Augustyniak$^\textrm{{\footnotesize\hyperlink{hl:warsaw}{23}}}$,
C.~D.~R.~Azevedo$^\textrm{{\footnotesize\hyperlink{hl:aveiro}{26}}}$\orcidlink{0000-0002-0012-9918},
B.~Badelek$^\textrm{{\footnotesize\hyperlink{hl:warsawu}{25}}}$\orcidlink{0000-0002-4082-1466},
R.~Beck$^\textrm{{\footnotesize\hyperlink{hl:bonniskp}{8}}}$,
J.~Beckers$^\textrm{{\footnotesize\hyperlink{hl:munichtu}{12}}}$\orcidlink{0009-0009-7186-255X},
Y.~Bedfer$^\textrm{{\footnotesize\hyperlink{hl:saclay}{6}}}$\orcidlink{0000-0002-5198-1852},
J.~Bernhard$^\textrm{{\footnotesize\hyperlink{hl:cern}{30}}}$\orcidlink{0000-0001-9256-971X},
F.~Bradamante$^\textrm{{\footnotesize\hyperlink{hl:triest_i}{17}}}$\orcidlink{0000-0001-6136-376X},
A.~Bressan$^\textrm{{\footnotesize\hyperlink{hl:triest_u}{18},\hyperlink{hl:triest_i}{17}, \hyperlink{hl:*}{*}}}$\orcidlink{0000-0002-3718-6377},
W.-C.~Chang$^\textrm{{\footnotesize\hyperlink{hl:taipei}{31}}}$\orcidlink{0000-0002-1695-7830},
C.~Chatterjee$^\textrm{{\footnotesize\hyperlink{hl:triest_i}{17},\hyperlink{hl:a}{a}}}$\orcidlink{0000-0001-7784-3792},
M.~Chiosso$^\textrm{{\footnotesize\hyperlink{hl:turin_u}{20},\hyperlink{hl:turin_i}{19}}}$\orcidlink{0000-0001-6994-8551},
S.-U.~Chung$^\textrm{{\footnotesize\hyperlink{hl:munichtu}{12},\hyperlink{hl:j}{j},\hyperlink{hl:j1}{j1}}}$,
A.~Cicuttin$^\textrm{{\footnotesize\hyperlink{hl:triest_i}{17},\hyperlink{hl:triest_a}{16}}}$\orcidlink{0000-0002-3645-9791},
M.~L.~Crespo$^\textrm{{\footnotesize\hyperlink{hl:triest_i}{17},\hyperlink{hl:triest_a}{16}}}$\orcidlink{0000-0002-5483-3388}
% ...

\vspace{10pt}
\hypertarget{hl:aanl}{$^\textrm{{\footnotesize 1}}$\footnotesize~A.I. Alikhanyan National Science Laboratory, 2 Alikhanyan Br. Street, 0036, Yerevan, Armenia$^\textrm{{\tiny\hyperlink{hl:A}{A}}}$\\}
\hypertarget{hl:brno}{$^\textrm{{\footnotesize 2}}$\footnotesize~Institute of Scientific Instruments of the CAS, 61264 Brno, Czech Republic$^\textrm{{\tiny\hyperlink{hl:B}{B}}}$\\}
\hypertarget{hl:liberec}{$^\textrm{{\footnotesize 3}}$\footnotesize~Technical University in Liberec, 46117 Liberec, Czech Republic$^\textrm{{\tiny\hyperlink{hl:B}{B}}}$\\}
\hypertarget{hl:praguectu}{$^\textrm{{\footnotesize 4}}$\footnotesize~Czech Technical University in Prague, 16636 Prague, Czech Republic$^\textrm{{\tiny\hyperlink{hl:B}{B}}}$\\}
\hypertarget{hl:praguecu}{$^\textrm{{\footnotesize 5}}$\footnotesize~Charles University, Faculty of Mathematics and Physics, 12116 Prague, Czech Republic$^\textrm{{\tiny\hyperlink{hl:B}{B}}}$\\}
\hypertarget{hl:saclay}{$^\textrm{{\footnotesize 6}}$\footnotesize~IRFU, CEA, Universit\'e Paris-Saclay, 91191 Gif-sur-Yvette, France\\}
\hypertarget{hl:bochum}{$^\textrm{{\footnotesize 7}}$\footnotesize~Universit\"at Bochum, Institut f\"ur Experimentalphysik, 44780 Bochum, Germany$^\textrm{{\tiny\hyperlink{hl:C}{C}}}$\\}
\hypertarget{hl:bonniskp}{$^\textrm{{\footnotesize 8}}$\footnotesize~Universit\"at Bonn, Helmholtz-Institut f\"ur  Strahlen- und Kernphysik, 53115 Bonn, Germany$^\textrm{{\tiny\hyperlink{hl:C}{C}}}$\\}
\hypertarget{hl:bonnpi}{$^\textrm{{\footnotesize 9}}$\footnotesize~Universit\"at Bonn, Physikalisches Institut, 53115 Bonn, Germany$^\textrm{{\tiny\hyperlink{hl:C}{C}}}$\\}
\hypertarget{hl:freiburg}{$^\textrm{{\footnotesize 10}}$\footnotesize~Universit\"at Freiburg, Physikalisches Institut, 79104 Freiburg, Germany$^\textrm{{\tiny\hyperlink{hl:C}{C}}}$\\}
\hypertarget{hl:mainz}{$^\textrm{{\footnotesize 11}}$\footnotesize~Universit\"at Mainz, Institut f\"ur Kernphysik, 55099 Mainz, Germany$^\textrm{{\tiny\hyperlink{hl:C}{C}}}$\\}
\hypertarget{hl:munichtu}{$^\textrm{{\footnotesize 12}}$\footnotesize~Technische Universit\"at M\"unchen, Physik Dept., 85748 Garching, Germany$^\textrm{{\tiny\hyperlink{hl:C}{C}}}$\\}
\hypertarget{hl:munichuni}{$^\textrm{{\footnotesize 13}}$\footnotesize~Ludwig-Maximilians-Universit\"at, 80539 M\"unchen, Germany\\}
\hypertarget{hl:calcutta}{$^\textrm{{\footnotesize 14}}$\footnotesize~Matrivani Institute of Experimental Research \& Education, Calcutta-700 030, India$^\textrm{{\tiny\hyperlink{hl:D}{D}}}$\\}
\hypertarget{hl:telaviv}{$^\textrm{{\footnotesize 15}}$\footnotesize~Tel Aviv University, School of Physics and Astronomy, 69978 Tel Aviv, Israel$^\textrm{{\tiny\hyperlink{hl:E}{E}}}$\\}
\hypertarget{hl:triest_a}{$^\textrm{{\footnotesize 16}}$\footnotesize~Abdus Salam ICTP, 34151 Trieste, Italy\\}
\hypertarget{hl:triest_i}{$^\textrm{{\footnotesize 17}}$\footnotesize~Trieste Section of INFN, 34127 Trieste, Italy\\}
\hypertarget{hl:triest_u}{$^\textrm{{\footnotesize 18}}$\footnotesize~University of Trieste, Dept.\ of Physics, 34127 Trieste, Italy\\}
\hypertarget{hl:turin_i}{$^\textrm{{\footnotesize 19}}$\footnotesize~Torino Section of INFN, 10125 Torino, Italy\\}
\hypertarget{hl:turin_u}{$^\textrm{{\footnotesize 20}}$\footnotesize~University of Torino, Dept.\ of Physics, 10125 Torino, Italy\\}
\hypertarget{hl:miyazaki}{$^\textrm{{\footnotesize 21}}$\footnotesize~University of Miyazaki, Miyazaki 889-2192, Japan$^\textrm{{\tiny\hyperlink{hl:F}{F}}}$\\}
\hypertarget{hl:yamagata}{$^\textrm{{\footnotesize 22}}$\footnotesize~Yamagata University, Yamagata 992-8510, Japan$^\textrm{{\tiny\hyperlink{hl:F}{F}}}$\\}
\hypertarget{hl:warsaw}{$^\textrm{{\footnotesize 23}}$\footnotesize~National Centre for Nuclear Research, 02-093 Warsaw, Poland$^\textrm{{\tiny\hyperlink{hl:G}{G}}}$\\}
\hypertarget{hl:warsawtu}{$^\textrm{{\footnotesize 24}}$\footnotesize~Warsaw University of Technology, Institute of Radioelectronics, 00-665 Warsaw, Poland$^\textrm{{\tiny\hyperlink{hl:G}{G}}}$\\}
\hypertarget{hl:warsawu}{$^\textrm{{\footnotesize 25}}$\footnotesize~University of Warsaw, Faculty of Physics, 02-093 Warsaw, Poland$^\textrm{{\tiny\hyperlink{hl:G}{G}}}$\\}
\hypertarget{hl:aveiro}{$^\textrm{{\footnotesize 26}}$\footnotesize~University of Aveiro, I3N, Dept. of Physics, 3810-193 Aveiro, Portugal$^\textrm{{\tiny\hyperlink{hl:H}{H}}}$\\}
\hypertarget{hl:lisbon}{$^\textrm{{\footnotesize 27}}$\footnotesize~LIP, 1649-003 Lisbon, Portugal$^\textrm{{\tiny\hyperlink{hl:H}{H}}}$\\}
\hypertarget{hl:dubna}{$^\textrm{{\footnotesize 28}}$\footnotesize~Affiliated with an international laboratory covered by a cooperation agreement with CERN\\}
\hypertarget{hl:russia}{$^\textrm{{\footnotesize 29}}$\footnotesize~Affiliated with an institute formerly covered by a cooperation agreement with CERN\\}
\hypertarget{hl:cern}{$^\textrm{{\footnotesize 30}}$\footnotesize~CERN, 1211 Geneva 23, Switzerland\\}
\hypertarget{hl:taipei}{$^\textrm{{\footnotesize 31}}$\footnotesize~Academia Sinica, Institute of Physics, Taipei 11529, Taiwan$^\textrm{{\tiny\hyperlink{hl:I}{I}}}$\\}
\hypertarget{hl:taipeincu}{$^\textrm{{\footnotesize 32}}$\footnotesize~Center for High Energy and High Field Physics and Dept.\ of Physics, National Central University, 300 Zhongda Rd., Zhongli 320317, Taiwan$^\textrm{{\tiny\hyperlink{hl:I}{I}}}$\\}
\hypertarget{hl:illinois}{$^\textrm{{\footnotesize 33}}$\footnotesize~University of Illinois at Urbana-Champaign, Dept.\ of Physics, Urbana, IL 61801-3080, USA$^\textrm{{\tiny\hyperlink{hl:J}{J}}}$\\}

\vspace{10pt}
\hypertarget{hl:*}{$^\textrm{{\footnotesize *}}$\footnotesize~Corresponding author\\}
\hypertarget{hl:a}{$^\textrm{{\footnotesize a}}$\footnotesize~Supported by the European Union’s Horizon 2020 research and innovation programme under grant agreement STRONG–2020 - No 824093\\}
\hypertarget{hl:b}{$^\textrm{{\footnotesize b}}$\footnotesize~Retired from Ludwig-Maximilians-Universit\"at, 80539 M\"unchen, Germany\\}
\hypertarget{hl:b1}{$^\textrm{{\footnotesize b1}}$\footnotesize~Supported by the DFG cluster of excellence `Origin and Structure of the Universe' (www.universe-cluster.de) (Germany)\\}
\hypertarget{hl:c}{$^\textrm{{\footnotesize c}}$\footnotesize~Also at ORIGINS Excellence Cluster, 85748 Garching, Germany\\}
\hypertarget{hl:d}{$^\textrm{{\footnotesize d}}$\footnotesize~Also at Institut f\"ur Theoretische Physik, Universit\"at T\"ubingen, 72076 T\"ubingen, Germany\\}
\hypertarget{hl:e}{$^\textrm{{\footnotesize e}}$\footnotesize~Supported by the Max Planck Institute for Physics, 85748 Garching, Germany\\}
\hypertarget{hl:f}{$^\textrm{{\footnotesize f}}$\footnotesize~Present address: NISER, Centre for Medical and Radiation Physics, Bubaneswar, India\\}
\hypertarget{hl:g}{$^\textrm{{\footnotesize g}}$\footnotesize~Also at University of Eastern Piedmont, 15100 Alessandria, Italy\\}
\hypertarget{hl:g1}{$^\textrm{{\footnotesize g1}}$\footnotesize~Supported by the Funds for Research 2019-22 of the University of Eastern Piedmont\\}
\hypertarget{hl:h}{$^\textrm{{\footnotesize h}}$\footnotesize~Also at Chubu University, Kasugai, Aichi 487-8501, Japan\\}
\hypertarget{hl:i}{$^\textrm{{\footnotesize i}}$\footnotesize~Also at KEK, 1-1 Oho, Tsukuba, Ibaraki 305-0801, Japan\\}
\hypertarget{hl:j}{$^\textrm{{\footnotesize j}}$\footnotesize~Also at Dept.\ of Physics, Pusan National University, Busan 609-735, Republic of Korea\\}
\hypertarget{hl:j1}{$^\textrm{{\footnotesize j1}}$\footnotesize~Also at Physics Dept., Brookhaven National Laboratory, Upton, NY 11973, USA\\}
\hypertarget{hl:k}{$^\textrm{{\footnotesize k}}$\footnotesize~Also at Fairmont State University, Department of Natural Sciences, 1201 Locust Ave, Fairmont, West Virginia 26554, USA\\}
\hypertarget{hl:l}{$^\textrm{{\footnotesize l}}$\footnotesize~Also at Dept.\ of Physics, National Kaohsiung Normal University, Kaohsiung County 824, Taiwan\\}
\hypertarget{hl:m}{$^\textrm{{\footnotesize m}}$\footnotesize~Also at RIKEN Nishina Center for Accelerator-Based Science, Wako, Saitama 351-0198, Japan\\}
\hypertarget{hl:n}{$^\textrm{{\footnotesize n}}$\footnotesize~Also at INFN TIFPA, 38123 Trento, Italy\\}
\hypertarget{hl:o}{$^\textrm{{\footnotesize o}}$\footnotesize~Also at Università di Trento, 38123 Trento, Italy\\}

\hypertarget{hl:$\dagger$}{$^\textrm{{\footnotesize $\dagger$}}$\footnotesize~Deceased\\}

\vspace{10pt}
\hypertarget{hl:A}{$^\textrm{{\tiny A}}$\footnotesize~Supported by the Higher Education and Science Committee of the Republic of Armenia (Armenia)\\}
\hypertarget{hl:B}{$^\textrm{{\tiny B}}$\footnotesize~Supported by MEYS, Grants LM2023040, LM2018104, LTT17018 and GAUK60121, CZ.02.01.01/00/22\_008/0004632 "FORTE", co-funded by the EU and Charles University Grant PRIMUS/22/SCI/017 (Czech Republic)\\}
\hypertarget{hl:C}{$^\textrm{{\tiny C}}$\footnotesize~Supported by BMBF - Bundesministerium f\"ur Bildung und Forschung (Germany)\\}
\hypertarget{hl:D}{$^\textrm{{\tiny D}}$\footnotesize~Supported by B. Sen fund (India)\\}
\hypertarget{hl:E}{$^\textrm{{\tiny E}}$\footnotesize~Supported by the Israel Academy of Sciences and Humanities (Israel)\\}
\hypertarget{hl:F}{$^\textrm{{\tiny F}}$\footnotesize~Supported by MEXT and JSPS, Grants 18002006, 20540299, 18540281 and 26247032, the Daiko and Yamada Foundations (Japan)\\}
\hypertarget{hl:G}{$^\textrm{{\tiny G}}$\footnotesize~Supported by NCN, Grant 2020/37/B/ST2/01547 (Poland)\\}
\hypertarget{hl:H}{$^\textrm{{\tiny H}}$\footnotesize~Supported by FCT, Grants DOI 10.54499/CERN/FIS-PAR/0022/2019 and DOI 10.54499/CERN/FIS-PAR/0016/2021 (Portugal)\\}
\hypertarget{hl:I}{$^\textrm{{\tiny I}}$\footnotesize~Supported by the Ministry of Science and Technology (Taiwan)\\}
\hypertarget{hl:J}{$^\textrm{{\tiny J}}$\footnotesize~Supported by the National Science Foundation, Grant no. PHY-1506416 (USA)\\}

\end{flushleft}

\clearpage
}

\maketitle

\section{Introduction}
The semi-inclusive measurement of hadron production
in deep-inelastic lepton–nucleon scattering (SIDIS),
l + N $\rightarrow$ l$'$ + h + X, is an important method for studying the
structure and formation of hadrons, widely used in theoretical and phenomenological studies to extract fragmentation functions. They describe the transition from quarks to observable hadrons. 
To reduce systematic uncertainties, experimental analyses often focus on the ratio of the semi-inclusive hadron production cross-section to the inclusive DIS cross-section, l + N $\rightarrow$ l$'$ + X. This ratio is referred to as multiplicity, see $e.g.$ Eq.~(1) in Ref. \cite{comp_pi}.

The COMPASS Collaboration has measured multiplicities of unidentified hadrons, pions and kaons using both isoscalar \cite{comp_pi, comp_K} and
proton \cite{comp_pi_prd} targets. 
Radiative corrections must be applied to both the hadron production and inclusive DIS measurements to convert the observed cross-sections to the one-photon exchange cross-sections,  $ rc = \sigma^{1 \gamma} / \sigma_{\rm observed}$.
The correction to the inclusive DIS cross-section $ rc_{\rm DIS}$ is well established,  see Refs. \cite{terad, rcbbkk}.
The correction to the hadron production cross-section ($i.e.$ SIDIS) $rc_{\rm had}$ is more involved.
The results published in Refs. \cite{comp_pi, comp_K} were obtained correcting the measured multiplicities by the factor $RC = rc_{\rm had} / rc_{\rm DIS}$, which directly quantifies the impact of QED radiative
correction effects on the multiplicities.

 In Ref. \cite{comp_pi} radiative corrections were estimated using the program TERAD, which is designed to compute inclusive DIS cross-sections \cite{terad, rcbbkk}. 
To estimate ${rc_{\rm had}}$, elastic and quasi-elastic contributions were subtracted from the observed cross-section. This method resulted in changes to the multiplicities of less than 2\% in the low-$x$, middle-$y$ region. 
Recognising that this was only an approximate method, COMPASS explicitly published the bin-by-bin values of the applied corrections, ${rc_{\rm DIS}}$ and ${ rc_{\rm had}}$, allowing future re-evaluations with improved methods. 
In Ref. \cite{comp_K}, which focused on kaon multiplicities, additional concerns were raised regarding the reliability of ${rc_{\rm had}}$.
As a result, a more conservative approach was adopted: the applied $rc_{\rm had}$ was taken as the average of the TERAD-based hadron- and the inclusive corrections. 
The associated uncertainty was increased accordingly, ensuring that it was always larger than half the difference between the TERAD-based hadron- and inclusive correction values.

To improve the reliability of hadron radiative corrections, the COMPASS Collaboration used dedicated Monte Carlo generators, in which also hadronic final states are simulated. 
At the time of the earlier publications, we attempted to use the program RADGEN \cite{radgen}. Unfortunately, this tool failed to consistently reproduce COMPASS data, which ultimately led to the continued reliance on TERAD-based estimates in Refs. \cite{comp_pi, comp_K}.

In recent years, however, the DJANGOH Monte Carlo generator (DJANGOH-MC) \cite{dj1, dj2}  has been successfully verified, $i.e.$ COMPASS achieved a satisfactory description of SIDIS data, which is demonstrated in Ref. \cite{comp_pi_prd}. 
This finally enabled the application of a more robust and theoretically consistent method for calculating radiative corrections. 
The values of $RC$ obtained from DJANGOH-MC differ significantly from those previously used, which is particularly important for data in Ref.~\cite{comp_pi}. 
In view of this, the COMPASS Collaboration has prepared the present addendum to update multiplicity results obtained using the isoscalar target and published in Refs. \cite{comp_pi, comp_K}.

\section{Radiative corrections}

\begin{figure}[!ht]
\centerline{\includegraphics[clip,width=1.0\textwidth]{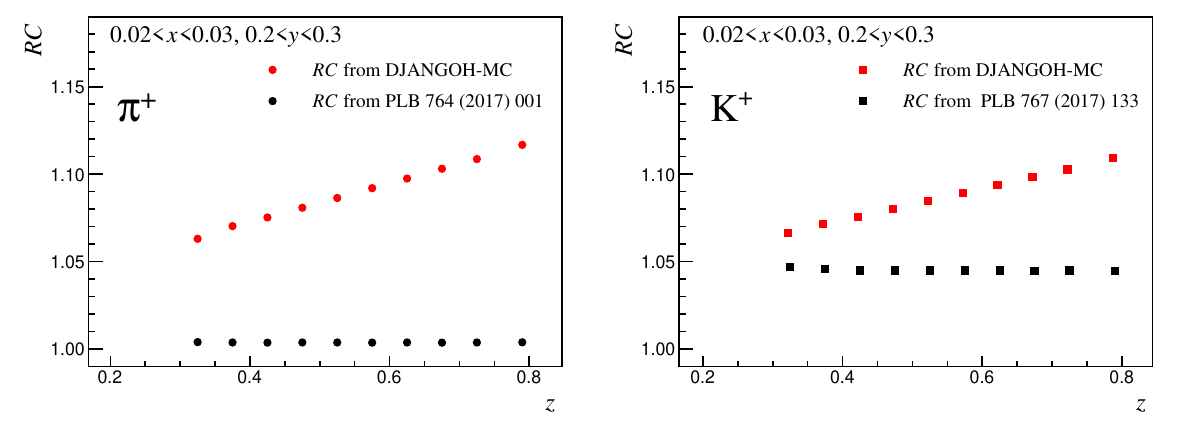}}
\caption{ Left panel: comparison of $RC$ factors for positive pions in a selected $(x,y)$ bin as a function of $z$  using DJANGOH-MC (red circles) and TERAD  from Ref. \cite{comp_pi} (black circles), 
right panel: comparison of $RC$ factors for positive kaons in a selected $(x,y)$ bin as a function of $z$ using DJANGOH-MC (red squares) and based on TERAD from Ref. \cite{comp_K} (black squares).}

\label{fig:rc1}
\end{figure}

The radiative correction factors $RC_{\rm DJANGOH}$, derived using DJANGOH-MC, were obtained using the two-step procedure described in Ref. \cite{comp_pi_prd}, in which $rc_{\rm DIS}$ is taken from TERAD and $rc_{\rm had}$ from DJANGOH.
In order to maintain consistency with the results reported in Ref. \cite{comp_pi_prd}, we present values of $RC_{\rm{DJANGOH}}$ rather than $rc_{\rm had}$. These values are presented in a three-dimensional binning of Bjorken variable $x$, lepton energy fraction carried by the virtual-photon $y$ and the fraction of the virtual-photon energy that is carried by the final-state hadron $z$.
 A comparison between $RC_{\rm DJANGOH}$ and the previously used correction factor $RC_{\rm TERAD}$ for pions \cite{comp_pi} and the bin $0.02 < x < 0.03$, $0.2 < y < 0.3$, as a function of $z$, is presented in the left panel of Fig.~\ref{fig:rc1}. Sizeable differences are observed.

The right panel of Fig.~\ref{fig:rc1} shows an analogous comparison for kaon multiplicities published in Ref. \cite{comp_K}. Here, the differences are smaller, primarily due to the conservative correction strategy employed for kaons, as described above. Nevertheless, the $z$-dependence of $RC_{\rm DJANGOH}$ remains evident and was not accounted for in the original analysis.

\begin{figure}[!ht]
\centerline{\includegraphics[clip,width=1.0\textwidth]{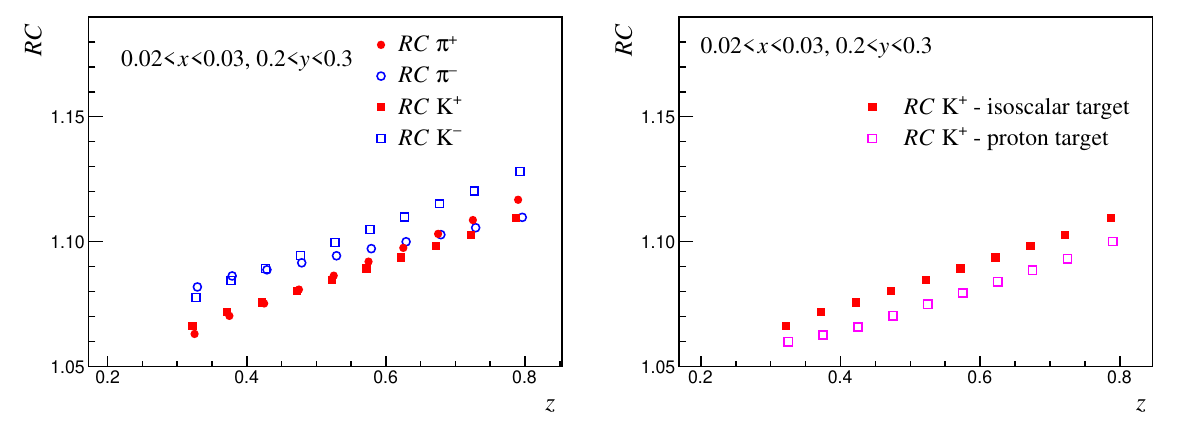}}
\caption{
Left panel: the values of $RC_{\rm DJANGOH}$ for $\pi^{\pm}$ and ${\rm K}^{\pm}$ in a selected $(x,y)$ bin as a function of $z$ (for clarity staggered horizontally), right panel: the values of $RC_{\rm DJANGOH}$ obtained for proton and isoscalar targets in a selected $(x,y)$ bin as a function of $z$.}
\label{fig:rc2}
\end{figure}

In the left panel of Fig.~\ref{fig:rc2}, we compare $RC_{\rm DJANGOH}$ for different hadron species and electric charges; only small differences are visible here. In the right panel of Fig.~\ref{fig:rc2}, a comparison of ${RC_{\rm DJANGOH}}$ between isoscalar and proton targets for positively charged kaons is shown; again, differences are small.

\section{New multiplicity results}

To obtain new isoscalar multiplicities,
radiative corrections are removed from the previous results \cite{comp_pi, comp_K} and the new  $RC_{\rm DJANGOH}$ are employed

\begin{equation}
    \frac{{\rm d} M_{\rm new}}{{\rm d}z} = \frac{ {\rm d} M_{\rm previous}}{{\rm d}z}\times \left(\frac{{rc_{\rm DIS}}}{{ rc_{\rm had}}}\right)_{\rm TERAD} \times {RC_{\rm DJANGOH}}.
\end{equation}
The statistical uncertainties are treated in the same way as the multiplicity values themselves, while the systematic uncertainties are kept unchanged.
As an illustrative example, in the kinematic region of $0.02 < x < 0.03$, $0.2 < y < 0.3$ and $0.30 < z < 0.35$ the published multiplicity for positively charged pions is:
$ {\rm d} M_{\rm previous}/{\rm d}z= 0.9427 \pm 0.0134_{\rm stat.} \pm 0.0472_{\rm syst.}$. 
The originally applied radiative correction factors are: ${rc_{\rm had}} = 0.9210$ 
and ${rc_{\rm DIS}} = 0.9173$, while
$RC_{\rm DJANGOH} = 1.0637$.
Applying these correction factors yields ${\rm d} M_{\rm new}/{\rm d}z= 0.9987 \pm 0.0142_{\rm stat.} \pm 0.0472_{\rm syst.}$. 
Note that in some bins in the corners of the kinematic region of the data published in Refs. \cite{comp_pi,comp_K} no $RC_{\rm DJANGOH}$
is obtained, thus these bins are omitted in the new tables.

\begin{figure}[!ht]
\centerline{\includegraphics[clip,width=1.0\textwidth]{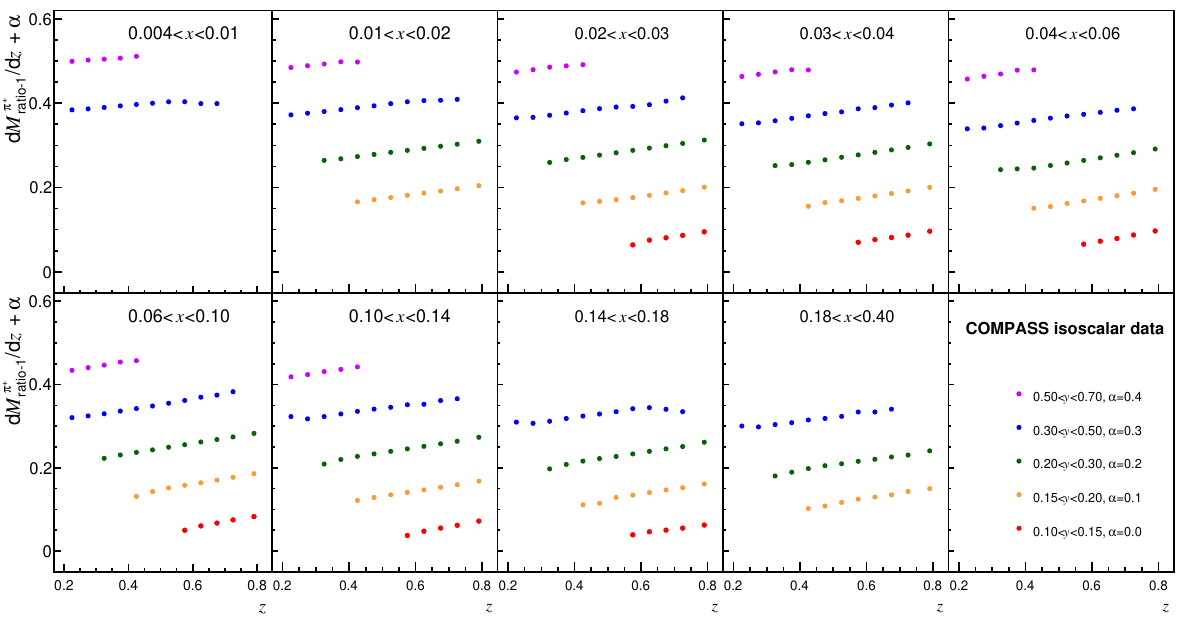}}
\caption{Values of ${\rm d} M^{\pi^{+}}_{\rm ratio-1} /{\rm d}z$ as a function of $z$ for nine bins of $x$ and five bins of $y$ (for clarity staggered vertically by~$\alpha$).}
\label{fig:rc3}
\end{figure}

\begin{figure}[!ht]
\centerline{\includegraphics[clip,width=1.0\textwidth]{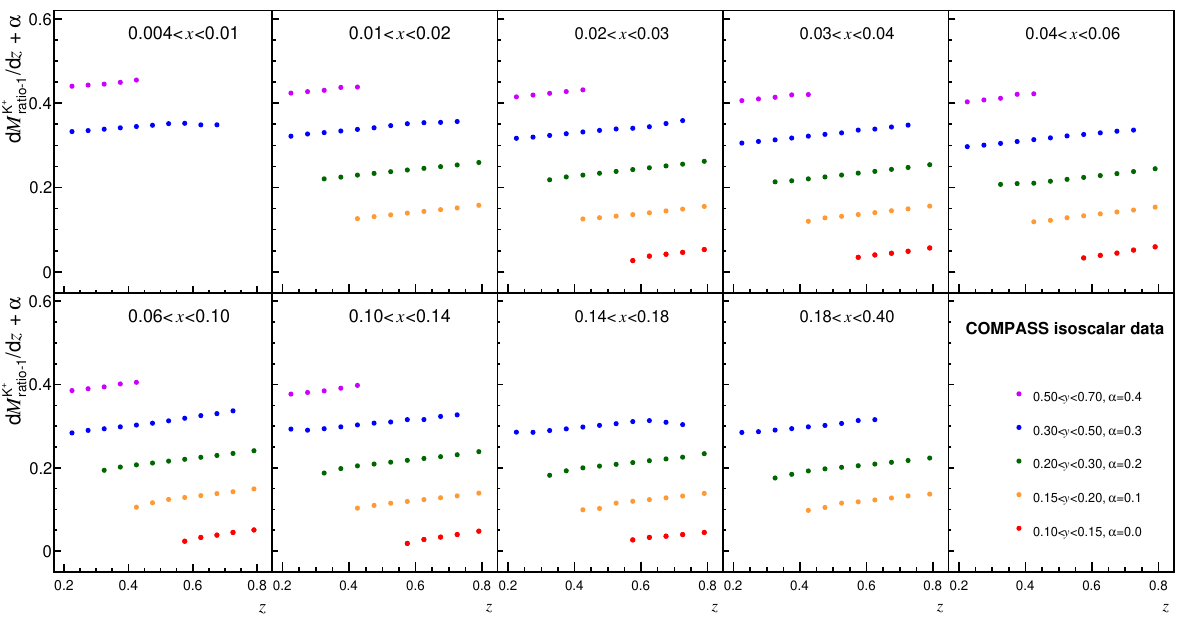}}
\caption{ Values of ${\rm d}M^{K^{+}}_{\rm ratio-1} /{\rm d}z$  as a function of  $z$ for nine bins of $x$ and five bins of $y$ (for clarity staggered vertically by~$\alpha$).}

\label{fig:rc4}
\end{figure}

In Figs. \ref{fig:rc3} and \ref{fig:rc4} we present   the ratio (${\rm d} M_{\rm new}/{\rm d}z) /({\rm d} M_{\rm previous}/{\rm d} z)-1$  for positive pions and kaons in bins of $(x,y)$ as a function of $z$. This ratio is denoted as
${\rm d}M^{\{\pi^{+}, \; K^{+}\}}_{\rm ratio-1} /{\rm d}z$ in Figs. \ref{fig:rc3} and \ref{fig:rc4}. The largest differences are present for low $x$, high $y$ and high $z$ and amount up to 12\%. As explained earlier, due to a different approach to $RC$ in the publications \cite{comp_pi} and \cite{comp_K}, the multiplicity changes for kaons are smaller than those for pions. Since multiplicities themselves decrease steeply with increasing $z$, these differences are not easily visible in typical plots such as Figs. 3$\div$8 of Ref. \cite{comp_pi}. Consequently, we chose to omit these figures in the present paper, as they do not lead to any new conclusions. 
The new multiplicity results as well as $RC_{\rm DJANGOH}$ are made available on the HEPData repository \cite{durham}.
These new results supersede the previous ones from Refs. \cite{comp_pi, comp_K}.  

\section{Summary}

In this Addendum, we present updated multiplicity values for an isoscalar target, published in Phys. Lett. B 764 (2017) 1 and Phys. Lett. B 767 (2017) 133. 
They are recalculated using radiative corrections obtained with the DJANGOH Monte Carlo generator. 
These radiative corrections lead to changes in the multiplicity values of up to 12\%. The present multiplicities supersede the previous ones.
Using them enables a consistent treatment of COMPASS isoscalar target multiplicity results with the proton multiplicities, recently published in Phys. Rev. D 112 (2025) 012002.

\section*{Acknowledgements}
We express our gratitude to H. Spiesberger for his valuable assistance in the use of the DJANGOH Monte Carlo generator, which significantly contributed to its successful
implementation for COMPASS. We gratefully acknowledge the support of the CERN management and staff and the skill and effort of
the technicians of our collaborating institutes. This work was made possible by the financial support of
our funding agencies.

\end{document}